\documentclass{emulateapj}

\slugcomment{Accepted for publication in the May 2005 issue of the PASP}

\shorttitle{FOBOS}
\shortauthors{Crane et al.}

\begin{document}

\title{The Fan Observatory Bench Optical Spectrograph (FOBOS)}

\author{Jeffrey D. Crane, Steven R. Majewski, Richard J. Patterson,
Michael F. Skrutskie, Elena Y. Adams, and Peter M. Frinchaboy}
\affil{Department of Astronomy, University of Virginia,
    Charlottesville, VA 22903}
\email{crane@ociw.edu, srm4n@virginia.edu, \\rjp0i@virginia.edu,
mfs4n@virginia.edu, eys5d@virginia.edu, \\pmf8b@virginia.edu}

%%%%%%%%%%%%%%%%%%%%%%%%%%%%%%% ABSTRACT %%%%%%%%%%%%%%%%%%%%%%%%%%%%%%%%

\begin{abstract}

The Fan Observatory Bench Optical Spectrograph (FOBOS) is intended for
single-object optical spectroscopy at moderate resolution 
($R\sim1500-3000$) using a fiber-fed, bench-mounted design to maintain 
stability. Whenever possible, the instrument uses off-the-shelf components 
to maintain a modest cost. FOBOS supports Galactic astronomy projects 
that require consistently well-measured ($\sim5$ km sec$^{-1}$) radial 
velocities for large numbers of broadly distributed and relatively bright 
($V\la14$) stars. The spectrograph provides wavelength coverage throughout 
the optical spectrum, although the instrument design was optimized for use
in the range $4700<\lambda<6700$~\AA. Test data indicate that the 
instrument is stable and capable of measuring radial velocities with
precision better than 3 km sec$^{-1}$ at a resolution of $R\sim1500$ 
with minimal calibration overhead.

\end{abstract}

\keywords{instrumentation: spectrographs, techniques: radial velocities}

%%%%%%%%%%%%%%%%%%%%%%%%%%%%% INTRODUCTION %%%%%%%%%%%%%%%%%%%%%%%%%%%%%%

\section{Introduction}

The Fan Observatory Bench Optical Spectrograph (FOBOS) addresses an 
increasing need for efficient moderate-resolution radial velocity surveys 
of bright stars distributed widely over the sky. As one example, the Grid 
Giant Star Survey (GGSS; Patterson et al. 2001) for NASA's \textit{Space 
Interferometry Mission (SIM)} requires spectroscopic study of several 
thousand bright ($V<$13.5) K giants evenly spaced about the sky, 
necessitating
large amounts of observing time on small aperture telescopes with 
single-object spectrographs. 
Few publicly available facilities with these capabilities exist, especially
in the northern hemisphere. In the southern hemiphere, the GGSS had used 
generous allocations of Las Campanas Observatory (LCO) Swope 1-meter time 
with the Modular Spectrograph, but this observing mode has since been 
decommissioned\footnote{Low resolution spectroscopy is, however, still 
available in the South on the CTIO 1.5-meter telescope.}. 
Moreover, the available facilities are predominantly slit spectrographs. 

Slit spectrographs suffer from both mechanical flexure and slit-centroiding 
errors that result in reduced velocity precision. Seeing can also produce
variable spectral resolution. Frequent lamp calibration can reduce flexure 
errors, but adds substantial overhead. A fiber-fed, bench-mounted
spectrograph can mitigate all of these problems, but with a penalty in 
system throughput. The benefits of stability and ease of use, however, can 
outweigh the decreased efficiency, particularly when targets are bright and
observing time is plentiful. Separating the spectrograph from the telescope
also protects the optics from environmental fluctuations (significant in
central Virginia) and
negates the requirement to fold the optical path into a compact unit 
mounted on the telescope.

The University of Virginia operates several telescopes at the Fan Mountain 
Observatory (FMO) in southern Albemarle County, Virginia. The 40 inch 
telescope is an f/13.5 astrometric reflector with a 15.0'' mm$^{-1}$ plate 
scale originally designed to continue the University's long-running 
astrometry program with the Leander McCormick Observatory $26\frac{1}{4}$-inch 
refractor after Charlottesville lights made work there difficult. 
Following construction in 1970, the 40-inch telescope collected
photographic plates for parallaxes. In the 1980's and 1990's, 
photographic plates gave way to electronic detectors for astrometry and 
more general photometric imaging, but the astrometry program switched focus 
to the southern hemisphere \citep{patterson91}.  The telescope has never 
been used for research quality 
spectroscopic work, although the site is perhaps 
better suited to that purpose (many nights each year are not photometric,
but still clear enough to allow spectroscopic observing). 

The proximity of the FMO site to the Department of Astronomy, the virtually
unlimited availability of observing time, and the good match of aperture 
size to several survey projects under consideration led to a proposal to 
build a fiber-fed optical spectrograph for the 40 inch telescope. This
proposal for an instrument of modest cost was also supported by the
\textit{SIM} project, which had interests in the GGSS expanding to the
northern hemisphere. The
instrument achieved first light in September 2003. In this paper, we 
discuss the philosophy, hardware
specifications and goals, overall design, construction, and
performance of FOBOS and give a brief overview of some of the planned 
science.

%%%%%%%%%%%%%%%%%%%%%%%%%%%%%%% GOALS %%%%%%%%%%%%%%%%%%%%%%%%%%%%%%%%%%

\section{Goals and Specifications}

FOBOS was designed primarily to provide northern hemisphere moderate
resolution spectroscopy for the GGSS. Since the southern hemisphere GGSS
spectroscopic campaign had already commenced on the 1-meter Swope 
telescope, the original goals for FOBOS were set simply to match the 
capabilities provided by the LCO Modular Spectrograph on that telescope. 
Specifically, the observed wavelength range was 4700 $<\lambda<$ 6700 \AA~
and the resolution aimed at $\sim$10 km s$^{-1}$ radial velocity accuracy. 
The purpose of the moderate resolution GGSS spectroscopy was to verify that
the photometrically selected giant candidates were in fact giant stars
(based on the Mg\textit{b}+MgH, H$\beta$, H$\alpha$, and NaD lines), and to
derive their metallicities, since low abundance giants are preferred for the
\textit{SIM} grid. The addition of 10 km sec$^{-1}$ velocities make the 
GGSS sample interesting for studies of Galactic dynamics.
GGSS targets are brighter than $V\sim$13.5, so a faint magnitude goal of 
$V=14$ was the primary specification, with the hope of achieving signal to 
noise ratios (S/N) of $\sim20$  at that magnitude in reasonable 
($10-15$ minute) exposures. Care was taken throughout the design process 
to make choices that would limit light losses and maximize efficiency. 

A limited hardware budget drove a design that maximized use of off-the-shelf
components. Access to an existing CCD and control electronics, a SITe
2048$\times$2048 CCD with $24\mu$m pixels, helped greatly to minimize cost,
but also imposed constraints on the end of the optical train since the CCD
was already fixed in a dewar with an unmodifiable back focus. At the front
end of the optical path, the telescope's f/13.5 optics guided the choice 
of optical fiber diameter, while the telescope back focus 
afforded by the
Cassegrain instrument mount restricted both the position of the fibers and
the options regarding target acquisition systems. 

%%%%%%%%%%%%%%%%%%%%%%%%%%%%%%% DESIGN %%%%%%%%%%%%%%%%%%%%%%%%%%%%%%%%%%

\section{Design}

Here we detail our solution to the described goals and constraints, 
a design that performs to specifications at a net expense of $<$\$35,000, 
less CCD and manpower costs, but including a new spectrograph enclosure 
within the observatory building.

\begin{figure}[htb]
   \epsscale{1.0}
   \begin{center}
   \plotone{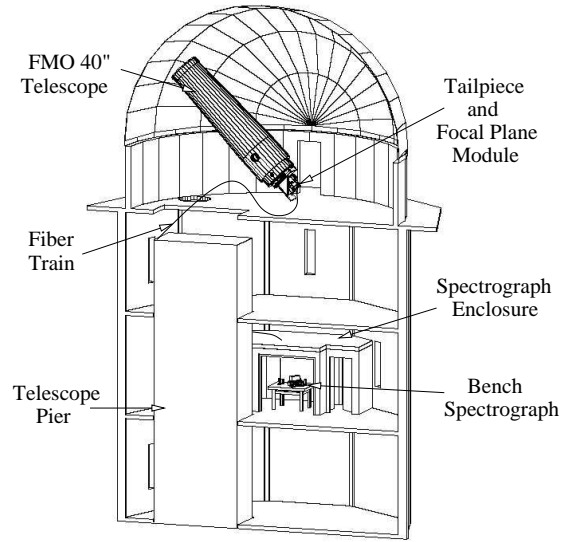}
   \end{center}
   \caption[fmobs]
   { \label{fig:fmobsall}
The complete FOBOS system on the FMO 40-inch telescope. The building 
and telescope pier are shown in cross-section. The fiber train runs
from the focal plane module through the telescope's polar axis (not shown)
and down the side of the telescope pier to the bench spectrograph
enclosure. The enclosure is shown with part of one wall removed.}
   \end{figure}

For the purpose of discussing instrument design and functionality, FOBOS 
can be separated into three main components. These are (1) the focal plane
module, which both provides a mechanism for aligning a target on an 
optical fiber and houses the calibration lamp assembly, (2) the fiber train, 
which transmits light from the telescope to the optical bench, and (3) the 
bench spectrograph itself, where spectral dispersion and data collection 
occur (Fig.~\ref{fig:fmobsall}). In addition to these components, an 
enclosure has been constructed to allow for more careful monitoring and 
control of the environment surrounding the bench spectrograph optics.

%-------------
%-------------

\subsection{The Focal Plane Module}

\begin{figure*}[htb]
   \epsscale{0.7}
   \begin{center}
   \plotone{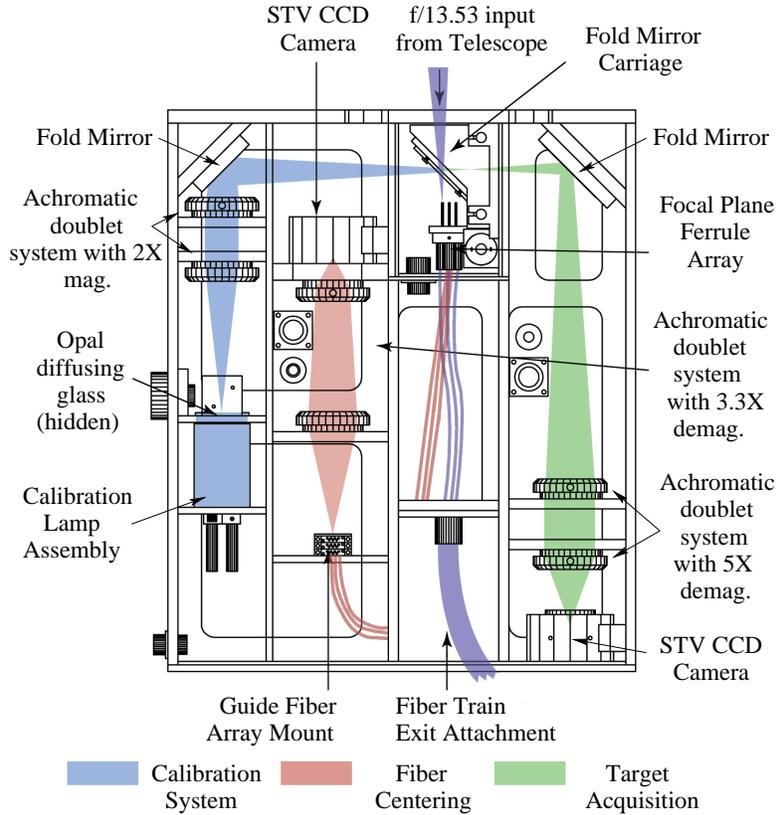}
   \end{center}
   \caption[fmobs]
   { \label{fig:fpmoddiag}
Diagram of the interior of the focal plane module.
}
   \end{figure*}

The focal plane module (Fig.~\ref{fig:fpmoddiag}) mounts to the base
of the telescope at the Cassegrain focus. A movable fold mirror
carriage within the module just above the telescope's focal plane 
enables three separate optical paths to fulfill three required functions.
%-------------
%-------------

\subsubsection{Focal Plane Viewing and Target Acquisition}

In the first fold mirror position, coarse target acquisition 
takes place.  The telescope image plane is demagnified five times by a 
system of two achromatic doublet lenses and viewed by a Santa Barbara 
Instrument Group (SBIG) model STV digital integrating video camera, 
yielding a 6.0$\times$4.4 arcmin field of view. The target star can be 
viewed, identified, and roughly positioned on the fiber by moving 
the telescope until the image of the star on a TV monitor is coincident 
with the known location of the fiber when the fold mirror is retracted. 
The fiber ends themselves cannot be seen with the STV camera 
because they lie beneath the fold mirror when the carriage is in this
position. It is also in this position 
that the telescope focus is adjusted. Relative to the movable fold mirror, 
the distances to the primary acquisition object plane and to the optical 
fiber array are equal, so that bringing into focus the image shown by the 
STV camera will also effectively focus the target on the fibers.

\subsubsection{Fiber Centering}

In the second fold mirror position, precise target placement on the fiber 
may be accomplished. Telescope light comes into focus in the plane of the 
focal plane fiber ferrules. Guide fibers (Section~\ref{fibertrain}) are 
monitored with a second SBIG STV camera while the alignment of the target 
star is fine-tuned. To image the array of guide fibers, a system of two 
achromatic doublet lenses is used to demagnify the object plane 3.3 
times. When alignment is achieved, the target's light is centered on the 
science fiber (Sec.~\ref{fibertrain}) and travels to the bench 
spectrograph. At this point, the auxiliary telescope autoguider (mounted
piggyback on the telescope tube) may be engaged if necessary (e.g., for 
fainter stars) and observing can begin. 

\subsubsection{Spectral Calibration}

In the final fold mirror position, the optical fibers may be illuminated by
calibration lamp light. Three wavelength calibration lamps (Neon, Argon, 
and Xenon) are available as well as a 10 watt Quartz-Tungsten-Halogen 
(QTH) lamp. These lamps are all independently controlled and illuminate 
an opal diffusing glass that is in turn imaged onto the plane of the fibers. 
A system of two achromatic doublets magnifies the opal glass two times and
focuses the calibration light at f/13.5 to match the f/ratio created by the
telescope during target observation, and thereby creates the same collimator
filling factor at the bench spectrograph.

\pagebreak
\subsection{The Fiber Train}
\label{fibertrain}

We are using step-index optical fiber with a 200 $\mu$m low OH silica core
from Polymicro Technologies (type STU). In the telescope's 15.0''/mm focal
plane, this fiber diameter corresponds to 3'' on the sky, roughly twice 
the estimated average seeing for the FMO site.

The main length of the fiber train consists of seven redundant, 85 foot
long ``science fibers'' that transmit light from the telescope to the bench 
spectrograph two floors below the dome room. Although this is a single-object 
spectrograph, we opted to include seven fibers held in a static array at 
the focal plane for three reasons. First, while one fiber is 
collecting stellar light, the ``empty'' fibers can be used 
to collect blank sky light for subtraction during data reduction. Using 
the average of several independent blank-sky spectra decreases the noise
contribution of the sky subtraction to the difference spectrum. Second, 
the extra fibers provide redundancy that may be invaluable in the event
that individual science fibers are broken. Finally, use of seven fibers 
allows the bundling of the full length in a close-pack hexagonal array, 
which provides extra rigidity and reduces the possibility of fiber 
tangling.

\begin{figure}[htb]
   \epsscale{1.0}
   \begin{center}
   \plotone{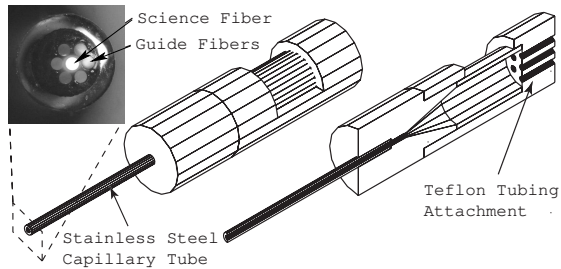}
   \end{center}
   \caption[fmobs]
   { \label{fig:fpferrule}
The focal plane fiber ferrule shown in isometric projection in full and
in cross-section. The teflon tube--encased fibers enter at the base where
the teflon tubes are epoxied to the aluminum. The fibers themselves are then
brought to a close-pack hexagonal array in the capillary tube extension. In
the photo, the central fiber has been back-illuminated. While the fibers in
the photo are themselves oriented well with respect to each other, the 
bundle is off-center within the capillary tube, a fact that is of no 
consequence to target acquisition.
}
   \end{figure}

In the telescope's focal plane, each science fiber is mounted in a ferrule
(Fig.~\ref{fig:fpferrule}) and surrounded closely by 6 short ``guide 
fibers,'' the back ends of which are viewed by a CCD camera, similar to the
spectrograph design employed in the Multi-Telescope Telescope \citep{barry}. 
During target acquisition and exposure, the guide fibers can be monitored 
for stellar light as a measure of the accuracy of alignment; when
each of the guide fibers is equally illuminated by light from
the wings of the stellar point spread function, the target star is 
centered on the array and therefore also the science fiber. In principle,
autoguiding of the telescope is possible via frame grabbing of the video
feed and evaluating relative fluxes in each guide fiber, as, e.g. with the
``Field Orientation Probes'' of the National Optical Astronomy 
Observatory's Hydra spectrograph \citep{barden}, but this has yet to be
implemented. In fact, we have found the telescope to track sufficiently 
well that manual guiding (using the telescope's hand paddle) is not an 
overly onerous task (and, of course, an advantage of fibers over slits is
that poor centering only translates to loss of light --- not a change in the
recorded equivalent ``slit function'').

%-------------
%-------------

After being cut to length, all fibers were inserted into black teflon tubes 
with 460~$\mu$m inner diameter
to provide a first layer of protection over the cladding as well as to prevent
cross-talk. The seven science fibers were then bundled together in a
close-pack hexagonal array for their full length and held in place using 
$\sim$1 cm lengths of shrink-wrap tubing spaced at two inch intervals. 
The spectrograph ends of the science fibers and the CCD ends of the guide 
fibers were terminated in 1/16'' outer diameter stainless steel capillary 
tubes with the fibers held in place using low-expansion epoxy. 
The fiber terminations 
in the telescope's focal plane are somewhat more complicated since each 
science fiber has to be closely surrounded by its six guide fibers
(Fig.~\ref{fig:fpferrule}). For these, the science fibers and surrounding
guide fibers were bundled together, coated with epoxy, and allowed to dry
before being inserted into capillary tube extensions with larger inside
diameters. The bundles were
then affixed inside the tubes using more epoxy.   
The completed science fiber train was fed through a flexible PVC conduit 
to provide some measure of protection against accidental crushing.

The ferrule ends and encased fibers were polished by hand using jigs that 
held the stainless steel capillary tubes against abrasive sheets. Six 
levels of aluminum oxide lapping film ranging from 30 to 0.3 $\mu$m grit 
sizes were used to refine the polish, following the procedure outlined 
by \citet{barry}.

Ideally, one would construct the complete fiber train and then test it to
assess the effects of focal ratio degradation (FRD) \textit{prior} to the
design of the bench spectrograph. However, the fiber work was very
time-consuming and this would have delayed progress. With the desire to
expedite the design process, we elected instead to evaluate a 2-meter test
fiber length prior to the construction of the actual fiber train. Using
the method described by \citet{carrasco}, it was estimated that with the
effects of FRD, approximately 90\% of the f/13.5 fiber input would exit
within an f/5 cone.

We chose to feed the fibers directly from the telescope without any
intervening fore-optics. As an alternative, however, one might consider the
benefits of using a focal reducer prior to the fiber entrance.
It is well established that the fractional FRD in a fiber improves as the
focal ratio of the input light is lowered (See Figure 8 of Carrasco \&
Parry 1993). Furthermore, the degree of FRD in a fiber is strongly
affected by the method in which it is terminated and by the quality of the
end polish. One might hope to minimize FRD such that for a small input focal
ratio, the majority of the light would exit within a similar cone angle.
The benefit of reducing the input focal ratio is that the fiber diameter
can then be reduced, resulting either in an increase of the spectral
resolution or in a relaxation of the grating angular dispersion
requirement and a subsequent effective increase in system throughput.
FOBOS as designed meets its resolution and efficiency requirements, so
fore-optics were avoided for the sake of cost and simplicity. However, the
benefits of fore-optics for other spectrograph designs should be carefully
considered.

\subsection{The Bench Spectrograph}
\label{bench}

\begin{figure*}[htb]
   \epsscale{0.65}
   \begin{center}
   \plotone{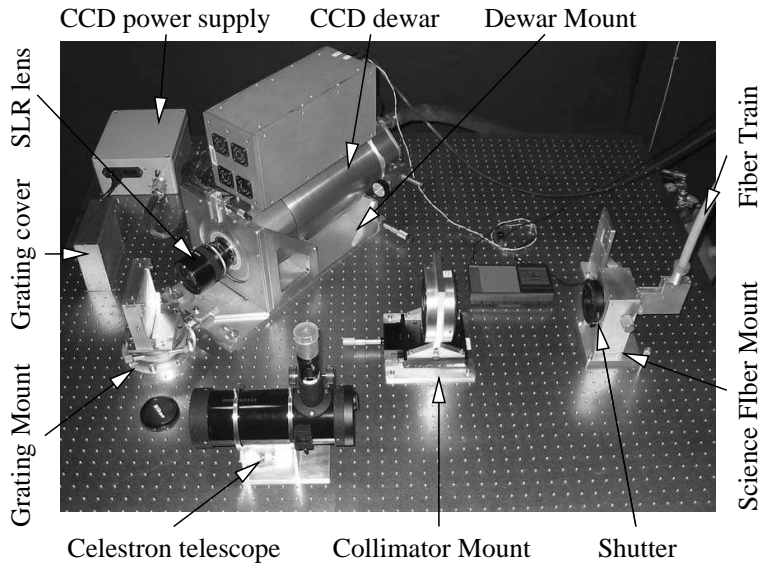}
   \end{center}
   \caption[fmobs]
   { \label{fig:bench}
The bench spectrograph. The fiber train attaches to the science fiber mount 
assembly on the right. After a curve in the fibers, light exits
horizontally. The CCD and camera rotate 
on a pivot arm (not visible in this view) to allow a range of 
collimator-grating-camera angles. The Celestron telescope is used to focus 
the collimator.
}
   \end{figure*}

The bench spectrograph (Fig.~\ref{fig:bench}) is mounted on a 4--inch 
thick optical table that sits on a pneumatically controlled vibration 
isolator. The seven science fibers are held in a linear array in 
the science fiber mount assembly, with the light exiting horizontally. 
An optional 100 $\mu$m slit mask can be inserted immediately in front of 
the fibers to increase resolution at the expense of a nearly 40\% 
throughput loss. By default, however, no mask is used, and the fiber ends
themselves define the effective entrance ``slit''. Following the slit, but
still within the science fiber mount, are two slots for optional filters. 
This location is easily accessible to the observer and allows for the use of
modestly sized (1 inch) filters. Due to space restrictions between the 
spectrograph camera optics and the CCD dewar, it was necessary to place 
the instrument's shutter at the front of the science fiber mount.

%-------------
%-------------

Following the results of the initial FRD testing, we restricted our search 
for a collimator to optics that were either f/5 naturally or that could be 
stopped to this f/ratio. Use of an on-axis paraboloidal reflector would have
resulted in light loss due to vignetting between the collimator and grating,
a negative consequence of using inexpensive reflective optics. This was
viewed as unacceptable given our desire to maximize instrument efficiency.
Cost considerations and limited off-the-shelf availability prevented us 
from choosing an off-axis paraboloidal reflector. Ultimately, then, we 
chose to use a lens for the collimator --- a 100 mm diameter, 350 mm focal
length achromatic doublet that could be stopped to f/5 with a 70
mm iris diaphragm mask on the collimated side of the beam. The collimator is
focused by manually adjusting the micrometer setting on the linear stage
that holds the lens mount, while simultaneously viewing the fiber and 
ferrule ends through a small telescope, focused at infinity, set in the 
collimated beam.

For the nominal GGSS spectral coverage, dispersion is provided by a 
1200 line/mm plane reflectance grating blazed at 21.1$^\circ$ for 
$\lambda=6000$\AA~ in first order. We chose this high groove density 
to give the necessary spectral range across our 2--inch detector in first 
order with the camera lens described below. Diffraction gratings with lower 
groove densities would not give us the desired GGSS spectral range with 
our collimator/camera combination. 
In this setup, the angle of incidence on the grating is smaller than the
angle of diffraction. This is an unconventional choice that helps to
reduce vignetting at the camera aperture since the anamorphic factor
reduces
the diameter of the diffracted monochromatic beam. Had we chosen to design
custom camera optics with a larger aperture, it would have been more
beneficial to choose a setup where the incident angle was larger than the
diffracted angle. In this case, the anamorphic factor would improve the
spectral resolution. Despite this unconventional setup on the FOBOS optical
bench, the required spectral resolution is achieved.
The grating is housed in an aluminum
structure that sits on a steel rotation stage operated to 1~arcminute
accuracy via adjustment of a manual drive with a vernier readout. The 
grating and grating 
housing can be easily removed from the rotation stage to allow use of other 
gratings from the inventory. Set screws in the grating housing allow
three-axis rotation of the grating itself so that the spatial alignment of 
spectra on the CCD can be fine-tuned.

Considering our demagnification requirements as well as the collimated beam
diameter, the ideal camera focusing optics would be somewhat faster than 
f/1.25 with a diameter a bit larger than 70~mm.  We were not able to find 
off-the-shelf optics that met these goals, but instead chose to use a 
Nikon telephoto 135mm f/2.0 SLR camera lens focused at infinity. 
The effective lens acceptance diameter is smaller than would be ideal, 
resulting in vignetting at the edges of the spectrum. The focal length 
is also longer than desired. However, it is still adequate to meet our
resolution goals. Because the optical design of such commercial lenses is 
proprietary, accurate and detailed modeling of the spectrograph optics could 
not be accomplished. Performance had to be estimated by assuming a paraxial
surface for the camera lens. 

The camera lens mounts directly to the front surface of the CCD dewar mount.
Although the back focal length of SLR lenses
are well established, the back focus of the camera lens must be
adjusted slightly for very red and very blue spectral coverage since SLR
lenses are not typically designed to be achromatic over a large range of
wavelengths.

The dewar mount rides on a steel alignment arm that pivots directly below
the center of the front surface of the grating. This swing arm keeps the
camera aligned with the grating and allows adjustment of the
collimator-grating-camera angle over a wide range. Dewar rotation about the
normal to the optical table may be accomplished by adjustment of a
micrometer setting and is useful for achieving optimal polychromatic focus
when tilt is present in the spectral focal plane. Dewar rotation about the
optical axis may be accomplished by adjusting a second micrometer and is
useful for achieving optimal alignment of spectra along CCD pixel rows.

Due to budgetary constraints, the spectrograph had to be designed to work
with one of FMO's existing research CCDs --- a SITe 2048$\times$2048 array
with 24 $\mu$m square pixels. With 2000 \AA~ of wavelength coverage, the best
possible spectral resolution that can be achieved is $\sim$2 \AA~ at the
Nyquist limit. In order to achieve this, the 200~$\mu$m optical fiber core
needed to be demagnified by the spectrograph optics by about a factor of 4.
Assuming the ability to centroid a single resolution element by a factor
of 20 (an extremely conservative assumption, as evidenced by our ability to
consistently attain considerably better effective centroiding using our 
own cross-correlation software), velocity resolutions
of $\sim$5 km s$^{-1}$ would be attainable if the Nyquist limit were
reached. However, the combination of camera and collimator optics chosen 
give a spectral resolution of $\sim$4 \AA~ after anamorphic magnification 
is considered. Although this meets the velocity resolution goals, an
optional 100~$\mu$m slit mask may be used just in front of the science 
fiber output in order to double the resolution to the Nyquist limit 
at the expense of a nearly 40\% throughput loss. The slit is thus 
useful for maximizing resolution capabilities, but is recommended only for 
brighter sources.

The specifications of the various bench optics are summarized in
Table~\ref{table:benchspecs}.

%-------------
   \begin{table}[htb]
   \centering
   \begin{tabular}{|ll|}
   \hline
   Parameter \hfill & Value\\
   \hline \hline
   Optical fiber core diameter \dotfill & 200 $\mu$m\\
   Collimator focal length \dotfill & 350 mm\\
   Collimator effective f/ratio \dotfill & f/5\\
   Grating angle of incidence \dotfill & -2.7$^\circ$\\
   Grating groove density \dotfill & 1200 lines/mm\\
   Grating blaze angle \dotfill & 21.1$^\circ$\\
   Grating blaze wavelength \dotfill & $\sim6000$~\AA\\
   Central diffraction angle \dotfill & 47.9$^\circ$\\
   Total collimator--camera angle \dotfill & 50.6$^\circ$\\
   Camera effective focal length \dotfill & 135 mm\\
   Camera effective f/ratio \dotfill & f/2\\
   Spatial demagnification \dotfill & 2.59$\times$\\
   Spectral demagnification \dotfill & 1.75$\times$\\
   Detector dimensions \dotfill & 2048$\times$2048\\
   Detector pixel width \dotfill & 24~$\mu$m\\
   Linear dispersion \dotfill & 1.0~\AA/pixel\\
   \hline
   \end{tabular}
   \caption{\label{table:benchspecs} Bench spectrograph optical component
   specification summary. These numbers are applicable to the FOBOS 
   configuration used for the GGSS.}
   \end{table}
%-------------

\subsection{The Spectrograph Enclosure}

To achieve the highest possible consistency
between observations, it is preferable to maintain a steady, controlled
environment around the optical table. The air should be still and at a
constant temperature throughout the night. Moreover, humidity in Virginia
can be a serious problem; condensation can easily develop on optical
surfaces if the dew point is not monitored and controlled. Finally,
because source light levels are so low --- especially after passing
through a fiber system --- sources of extraneous background light need to 
be eliminated. For these reasons, a special enclosure was constructed 
for the bench spectrograph on the second floor of the observatory, two 
floors below the telescope. 

The enclosure's walls and ceiling are six inches thick and filled with 
fiberglass insulation. A double door entryway leads to an
interior coated with flat black, non-fluorescent paint illuminated with 
incandescent lights. A window-mount air conditioner is set directly into 
one wall with a framed, well-sealed door that covers it at night. A 
dehumidifier keeps the humidity at safe levels. In addition, an air 
cleaner runs continuously during the day to minimize airborne dust. 
These devices are powered off prior to nighttime observing in order to let 
the air equilibrate and still.

An environmental monitoring and control system has been designed to 
maintain appropriate conditions inside the spectrograph enclosure. Two 
resistence temperature detectors monitor the concrete floor and air 
temperatures while a humidity sensor measures the air's water content. 
These sensors are connected to a National Instruments Field Point 
distributed I/O network that feeds data to a personal computer in the 
observatory control room. The PC is running a National Instruments 
LabView-based program that monitors conditions in the room and can
operate the environmental control devices as necessary. During the day, 
the room may be cooled to match the temperature of the concrete floor. 
During nighttime observing, the thermal mass of the floor maintains the 
air temperature reasonably well. 

%%%%%%%%%%%%%%%%%%%%%%%%%%%% DATA CALIBRATION %%%%%%%%%%%%%%%%%%%%%%%%%%%

\section{Data Calibration}

Data calibration is accomplished using a combination of internal comparison
lamp exposures and on-sky observations.

Each night, a set of bias frames is collected, combined, and 
subtracted from each data frame. Flat field images are created by placing 
an opal diffusing glass in one of the two filter slots in front of the 
fiber output in the bench spectrograph. The QTH lamp is used to illuminate 
the opal glass with bright, diffuse light. Since the opal glass is out of 
focus relative to the collimator, it acts as a good flat source useful for 
removing pixel-to-pixel variations in the spectral images. These frames are 
referred to as ``milky flats'', following the convention used with the Hydra
bench spectrograph at the Cerro Tololo Inter-American Observatory.

In most cases, multiple science fibers will be used to
collect background sky emission while the central science fiber collects 
light from the target star. In order to combine the multiple sky spectra
effectively, relative throughput differences between them need to be 
estimated and taken into account prior to subtraction from the stellar 
spectrum. This is best accomplished by observing the daytime or twilight 
sky. The uniform illumination over all fibers in the telescope's focal 
plane is very good for estimating fiber throughput variations. Once
throughput variations are removed, sky spectra are combined in wavelength
space after dispersion correction with reference to calibration lamp
exposures.

One-dimensional spectra are extracted from the images with reference to 
apertures determined by tracing the paths of spectra created by the QTH
lamp. A one-second exposure suffices to produce broadband continua 
bright enough to map out the traces of each fiber's spectrum across the 
detector. 

Once extracted to one dimension, spectra may be calibrated to determine the
wavelength-to-pixel mapping. Three wavelength calibration lamps are 
available: Neon, Argon, and Xenon. Each are controlled independently, 
a necessity since the Neon lamp is considerably brighter than the others. 
Typical exposure lengths for wavelength calibration are 60 seconds for Xenon
and Argon and 1 second for Neon. In that 
time, enough bright emission lines appear to span the wavelength ranges of 
interest so that good pixel/wavelength calibrations may be determined. 
It appears that the spectrograph is stable enough that wavelength calibration
spectra need only be collected once per night (see Section~\ref{stability}). 

%For our normal radial velocity programs, velocities of target stars are
%determined by cross-correlating spectra with those of radial velocity
%standard stars of similar spectral type. During a typical night, one might
%easily observe five to ten radial velocity standards in addition to scientific
%targets of interest. Finally, some observing programs may find it helpful to
%observe spectrophotometric standard stars in order to facilitate flux
%calibration.

%%%%%%%%%%%%%%%%%%%%%%%%%%%%% DATA REDUCTION %%%%%%%%%%%%%%%%%%%%%%%%%%%%

\section{Data Reduction}
\label{reduction}

Data are reduced using a combination of native and locally
written \textsc{iraf} \citep{tody} tasks as well as a radial velocity 
cross-correlation program written at the University of Virginia.

\begin{figure}[htb]
   \epsscale{1.1}
   \begin{center}
   \plotone{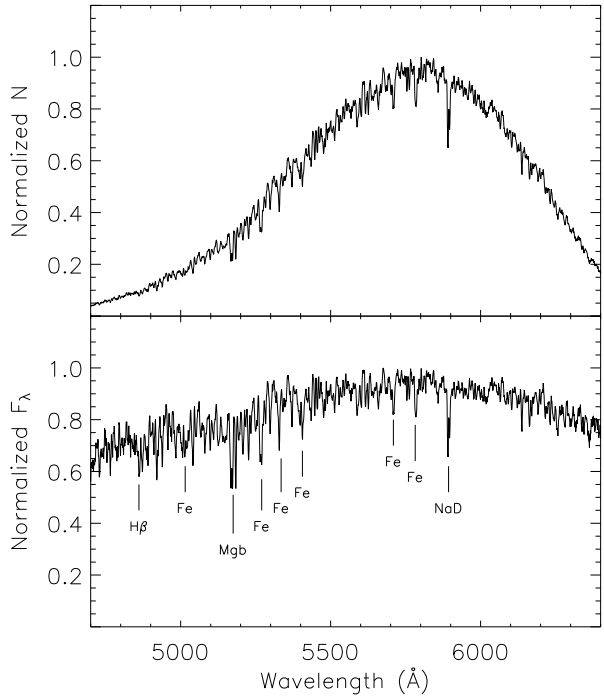}
   \end{center}
   \caption[spectrum]
   { \label{fig:spectrum}
FOBOS spectrum of HD~26162, a K1 giant with [Fe/H]$=-0.02$. The upper panel
shows normalized counts in the raw spectrum while the lower panel shows the
normalized, flux-calibrated spectrum. Many useful
absorption features are visible including H$\beta$ (4861\AA), the Mg{\it b}
triplet (5167-5184\AA), the Na~I doublet, and a multitude of Fe lines. 
In addition, the MgH band is present near 5150\AA, but is not obvious 
in the spectrum of this low surface gravity star.
}
   \end{figure}

An \textsc{iraf} package called \textsc{fobos} has been written to
streamline and partially automate data reduction. After median combination
of a set of bias frames, all data and calibration frames are 
bias-subtracted, overscan-corrected, and trimmed. A flat field image is 
then generated by combining the milky flat frames and dividing out the
large-scale image structure to leave only pixel-to-pixel variations. The 
flat field is then divided into the remaining frames. Next, QTH lamp and 
daytime sky exposures are extracted using the \textsc{iraf apall} routine.
The latter are used to estimate the fiber-to-fiber throughput differences, 
which are useful later during sky subtraction. If target observations have 
been split up into multiple, adjacent exposures (recommended to facilitate
removal of cosmic rays), these are then combined at this point. Finally, a 
new routine called \textsc{fobosext} is run that extracts the wavelength 
calibration and object images to 1-dimensional spectra, determines the 
wavelength-to-pixel mapping of the arc lamp exposures by interactively 
prompting the user to identify emission lines in the arc lamp spectrum, 
dispersion corrects the object spectra, and subtracts the background sky 
emission after combining spectra obtained from the blank sky fibers. A
sample spectrum is shown in Figure~\ref{fig:spectrum}.

Following these reductions, radial velocities may be estimated by
cross-correlating the extracted, wavelength-calibrated target spectra
against a set of radial velocity standard stars of similar spectral type.
This is executed using a \textsc{matlab} code written locally by M. Garvin, 
and following the general precepts for filtered cross correlation described 
by \citet{majewski04}. 
%This method has been consistently yielding better velocity results than one
%would normally expect using the more popular radial velocity measurement
%software (See Section~\ref{stability}).

%-------------
%-------------

%%%%%%%%%%%%%%%%%%%%%%%%%%%%%% PERFORMANCE %%%%%%%%%%%%%%%%%%%%%%%%%%%%%%

\section{Performance}

\subsection{Throughput}

\begin{figure}[htb]
   \epsscale{1.1}
   \begin{center}
   \plotone{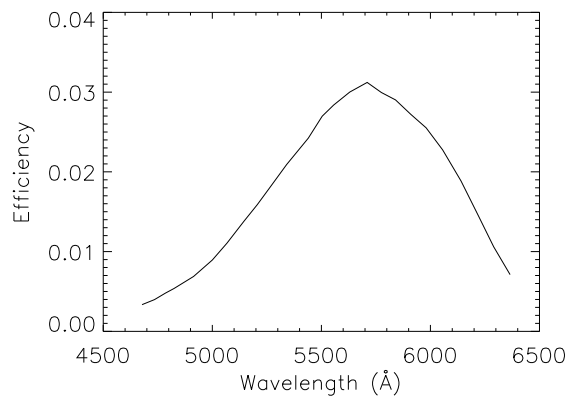}
   \end{center}
   \caption[efficiency]
   { \label{fig:efficiency}
Throughput vs. wavelength for the telescope + FOBOS system in the GGSS 
setup. The efficiency of FOBOS alone should be higher because we expect 
considerable contribution to the reduced efficiency (especially at shorter
wavelengths) by the telescope mirrors that had not been aluminized for 
several years at the time these data were collected.
}
   \end{figure}

The throughput of the FOBOS + telescope system was estimated by observing
the spectrophotometric flux standard star Feige 110 under photometric 
conditions on UT 2003 November 29. 
Following the standard CCD reductions and spectral extraction
described above, the accumulated counts in the observed spectrum were
compared to the expected number calculated from the Fiege 110 
spectrophotometric data given in the \textsc{iraf irscal} database. This
database gives the calibrated, monochromatic AB system magnitudes for the
standard star in a series of either 49~\AA ~or 98~\AA ~bins. These magnitudes
were converted to predicted counts at the telescope's entrance aperture
using the formula:

%-------------
\begin{eqnarray}
DN (\textnormal{ADU}) = 3.68\times10^{-20}\times10^{-0.4m_{AB}}\nonumber\\
  \times 10^{-0.4k_{\lambda}\times X}\times\frac{c\times10^{8}}
  {\lambda(\textnormal{\AA})^{2}}\times\frac{\lambda(\textnormal{cm})}
  {hc}\nonumber\\
  \times \pi(r_{prim}^{2}-r_{sec}^{2})\times\frac{Exp}{Gain}\times
  \Delta\lambda_{bin}(\textnormal{\AA})
\end{eqnarray}
%-------------

\noindent
where $m_{AB}$ and $k_{\lambda}$ are the calibrated magnitude and 
atmospheric extinction,
respectively, in the wavelength bin centered on $\lambda$ with width 
$\Delta\lambda_{bin}$, $X$ is the airmass of the observation, $c$ is the
speed of light in cm sec$^{-1}$, $h$ is Planck's constant in erg s,
$r_{prim}$ is the radius of the primary telescope aperture, $r_{sec}$ is the
radius of the telescope's secondary mirror cell, $Exp$ is the exposure time
of the observation, and $Gain$ is the CCD gain in electrons/ADU. The
resulting expected counts in each wavelength bin were then divided into the
observed counts in the same bin to produce the efficiency curve in
Figure~\ref{fig:efficiency}.

At the time that these efficiency data were collected, the telescope's
mirrors had not been aluminized for more than four years. Due to persistent
problems with humidity and condensation at the site combined with the
advanced age of the mirror coatings, we expect that the telescope's
efficiency has adversely affected the total system efficiency as
plotted in Figure~\ref{fig:efficiency}. In addition, seeing conditions were
somewhat worse than average for the FMO site, so that performance
here may be underestimated. Moreover, the blue response of the
mirror coatings has likely decayed more quickly than the red response.
Therefore, we expect that this efficiency estimate does not accurately
reflect the actual performance of the spectrograph itself. Following the
next scheduled re-aluminization of the telescope mirrors, a significant 
improvement in the overall throughput, especially at bluer wavelengths,
is expected. Overall efficiency will also improve with better seeing. Thus,
Figure~\ref{fig:efficiency} gives a lower limit to the performance of FOBOS.

%-------------
%-------------

\begin{figure}[htb]
   \epsscale{1.1}
   \begin{center}
   \plotone{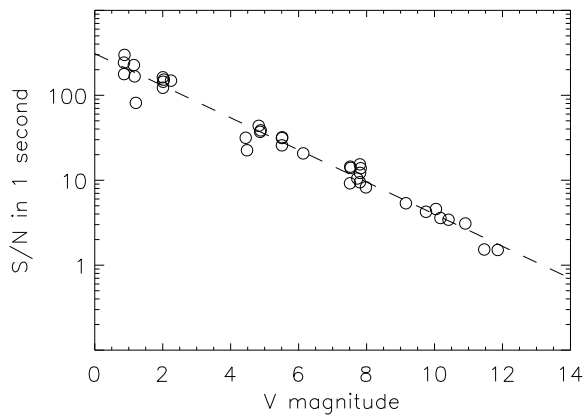}
   \end{center}
   \caption[snrrate]
   { \label{fig:snrrate}
Peak signal to noise ratio per pixel generated in one second at one airmass.
This represents a practical view of the instrument's sensitivity.
}
   \end{figure}

\begin{figure}[htb]
   \epsscale{1.1}
   \begin{center}
   \plotone{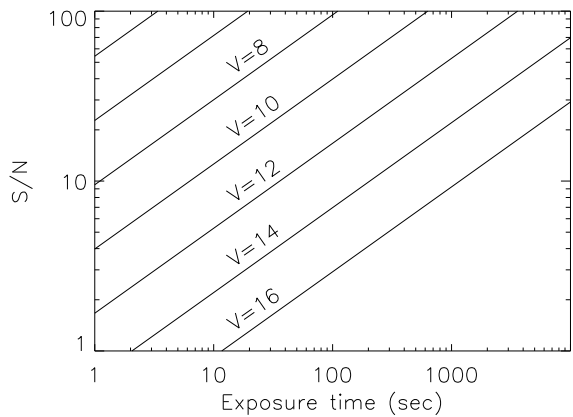}
   \end{center}
   \caption[exptime]
   { \label{fig:exptime}
Lower limit to the predicted signal to noise ratio vs. exposure time for 
different apparent magnitudes.
}
   \end{figure}

It is of more practical use to explore the instrument's sensitivity in terms
of the achievable signal-to-noise ratio (S/N) as a function of source 
magnitude.  A number of calibrating stars covering a range of magnitudes 
were observed on UT 2003 October 31 and UT 2003 November 29. All were 
reduced using the methods described in Section~\ref{reduction} and the 
resulting spectra were evaluated to determine their peak S/N. The results 
are plotted in Figure~\ref{fig:snrrate}, which shows the effective S/N that 
would be achieved for each star in one second of exposure (note, however,
that the actual exposures used to perform this evaluation were longer than
one second). 
A linear fit to the data was derived with equal weighting for all
37 points. Based on this fit, exposure times may be predicted for any 
source $V$ magnitude and target S/N (Fig.~\ref{fig:exptime}). 

%-------------
%-------------

%-------------
%-------------

\subsection{Velocity Measurement Accuracy and Stability}
\label{stability}

FOBOS was built, in part, to provide stable radial velocity measurements. 
On UT 2003 November 29, 26 spectra of radial velocity standards were 
collected, several of which were repeat observations of individual stars.
The data were then reduced using two different methods.

In the first method, each spectrum was extracted with an aperture referenced
to a QTH lamp exposure taken immediately after the object exposure. The
spectra were then wavelength-calibrated using arc lamp spectra collected
immediately after the adjacent QTH exposures. In the second method, all
spectra were extracted and wavelength-calibrated using only a single QTH and
arc lamp exposure pair taken at the beginning of the night.

Within each of these two groups, the spectrum of each star was
cross-correlated with all of the others, producing 25 individual measurements
of each star's radial velocity. Each star's 25 velocity measurements were 
then averaged and a standard deviation was calculated. The mean standard
deviations reported for each star's velocity measurements were 2.3 km
sec$^{-1}$ and 2.8 km sec$^{-1}$ for methods 1 and 2, respectively.
For bright ($V<$8) stars, this appears to be the limit of the instrument's
velocity precision for a single spectrum produced with the 1200 line/mm 
grating while observing without 
the optional aperture slit, and is well within the original design goals. 

\begin{figure}[htb]
   \epsscale{1.1}
   \begin{center}
   \plotone{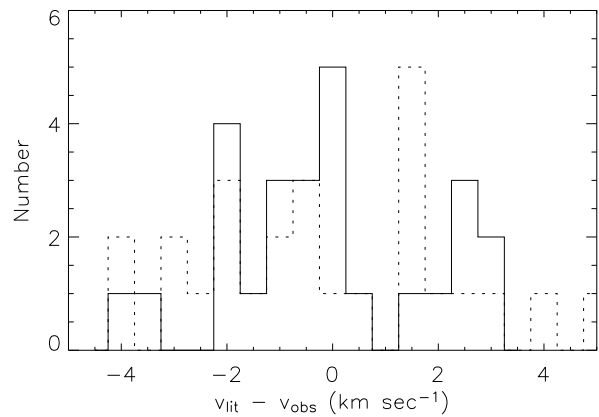}
   \end{center}
   \caption[stability]
   { \label{fig:stability}
Errors in velocity measurements for 26 radial velocity standard star 
observations reduced using two separate methods. The solid and dotted lines
correspond to the reductions using methods 1 and 2, respectively (See text
of Section~\ref{stability}). } \end{figure}

For spectra reduced using method 1, the mean velocity error (defined as 
$\overline{v}_{err} = N^{-1} \sum |v_{lit} - v_{obs}|$, where $v_{obs}$ is 
the measured velocity and $v_{lit}$ is the velocity taken from the 
Astronomical Almanac \citep{usno}) and dispersion about the mean was 
$\overline{v}_{err} = 1.5 \pm 1.2$ km sec$^{-1}$ while method 2 resulted in 
$\overline{v}_{err} = 2.0 \pm 1.3$ km sec$^{-1}$ (note that these numbers do
not take into account the uncertainties inherent in the published standard
velocities, some of which are as high as 0.5 km sec$^{-1}$).
Thus, the systematic errors are small. The distribution of 
measurement errors is shown in Figure~\ref{fig:stability}.
Based on these numbers, it appears that the results one gets from using only
a single set of calibration lamp exposures at the beginning of the night are
only marginally worse than the results obtained from taking many 
calibration frames
throughout the night. For observing programs that can tolerate the small
impact to velocity uncertainties, collecting only one set of calibration
frames will result in a significant improvement in observing efficiency.

%-------------
%-------------

%%%%%%%%%%%%%%%%%%%%%%%%%%%%%% FUTURE WORK %%%%%%%%%%%%%%%%%%%%%%%%%%%%%%%

\pagebreak
\section{Science and Future Work}

FOBOS was commissioned in September 2003 and is now being used for 
scientific research. Several projects are underway and planned for the 
immediate future. 

The northern hemisphere spectroscopic GGSS observing has already begun. 
As part of this program, several thousand candidate metal-poor K giants in 
the Galactic halo will be observed to further confirm their suitability 
as members of the \textit{SIM} Astrometric Grid, required to establish the
positional reference frame for that mission. In addition to providing 
necessary support for \textit{SIM}, however, the GGSS will be a unique
catalog of low-metallicity K giants suitable for exploring the structure of
the Galactic halo. GGSS observing in the southern hemisphere has already
yielded interesting, serendipitous results such as the discovery of
kinematic evidence for the presence of a tidal structure associated with 
the Sagittarius (Sgr) dwarf galaxy \citep{kundu}.

Work is also underway to measure the contribution of dark matter to the 
local Galactic disk mass density through the measurement of the density and
kinematics of a population of K giant tracers in the direction of the North
Galactic Pole \citep{crane04}. Spectroscopic observations and 
determinations of tracer star
velocities were in part executed using FOBOS during Spring 2004.

FOBOS is being used to characterize the occurrence of
emission lines within the complete B-star populations of an
age-selected sample of open clusters to a distance of 1~kpc
in a study of the evolutionary status of Be stars. Red spectra
including the H$\alpha$ 6563\AA~ and \ion{He}{1} 6678\AA~ lines can be
obtained for stars as faint as $V=13$, reaching the latest B
spectral types missing in previous studies and providing
positive identification of Be stars with weak emission that
are difficult to detect using narrow band photometry.

Finally, FOBOS will be used to measure the radial velocities of M giant
candidate members of both the Sgr tidal tail system \citep{majewski} and 
the Galactic Anticenter Stellar Stream \citep{yanny,crane} to further 
constrain their orbits and to gain insight into the nature of the Milky 
Way's gravitational potential.

%%%%%%%%%%%%%%%%%%%%%%%%%%%%% ACKNOWLEDGMENTS %%%%%%%%%%%%%%%%%%%%%%%%%%%

\acknowledgments

We thank James Barr and Charles Lam for lending their technical skills and
expertise, without which this project would have been considerably more
difficult. David McDavid, Howard Powell, Robert Rood, and Kiriaki Xilouris
each provided support at various stages of the project. We thank Megan 
Kohring for her assistance in constructing the fiber train. William Kunkel, 
Di Harmer, Robert Jones, Robert O'Connell, Ray Ohl, Neill Reid, and John 
Wilson provided insightful comments during useful discussions. We thank 
Matthew Garvin for providing software that aided our analyses. Finally, we
thank Daniel Fabricant and the anonymous referee for editorial comments that
helped to improve this paper.
This work has been supported by NASA/JPL grant 1222563 as part of
the NASA \textit{Space Interferometry Mission} and a Cottrell Scholar Award 
from The 
Research Corporation. This research was also partially supported by the 
The F.H. Levinson Fund of the Peninsula Community Foundation
and a David and Lucile Packard Foundation fellowship to
S.R.M. J.D.C. has been aided by a Grant-in-Aid of Research 
from the National Academy of Sciences, administered by Sigma Xi, The 
Scientific Research Society, and by an Aerospace Graduate Research 
Fellowship, administered by the Virginia Space Grant Consortium.

%%%%%%%%%%%%%%%%%%%%%%%%%%%%%%%% REFERENCES %%%%%%%%%%%%%%%%%%%%%%%%%%%%%%


\begin{thebibliography}{}

%\bibitem[(2004)]{usno} The Astronomical Almanac for the year 2003, 2001
%(Washington: USGPO)

\bibitem[Barden et al.(1993)]{barden} Barden, S.~C., Armandroff, T., 
Massey, P., Groves, L., Rudeen, A.~C., Vaughnn, D., \& 
Muller, G.\ 1993, in ASP Conf. Ser 37, Fiber Optics in Astronomy, II, 
ed. P.~M. Gray (San Francisco: ASP), 185 

\bibitem[Barry(1996)]{barry} Barry, D. J. 1996, Ph.D. Thesis, Georgia State
University

\bibitem[Carrasco \& Parry(1993)]{carrasco} Carrasco, E. \& Parry, I. R. 
1993, in ASP Conf. Ser 37, Fiber Optics in Astronomy, II,
ed. P.~M. Gray (San Francisco: ASP), 392

\bibitem[Crane et al.(2003)]{crane} Crane, J.~D., Majewski, 
S.~R., Rocha-Pinto, H.~J., Frinchaboy, P.~M., Skrutskie, M.~F., \& Law, 
D.~R. 2003, \apjl, 594, L119

\bibitem[Crane(2004)]{crane04} Crane, J.~D. 2004, Ph.D. Thesis, University
of Virginia

\bibitem[Kundu et al.(2002)]{kundu} Kundu, A.~et al. 2002, \apjl, 576, L125 

\bibitem[e.g., Majewski et al.(2003)]{majewski} 
Majewski, S.~R., Skrutskie, M.~F., Weinberg, M.~D., \& Ostheimer, J.~C. 
2003, \apj, 599, 1082 

\bibitem[Majewski et al.(2004)]{majewski04} Majewski, S.~R., et 
al.\ 2004, \aj, 128, 245 

\bibitem[e.g., Patterson \& Ianna(1991)]{patterson91} Patterson, R.~J., 
\& Ianna, P.~A.\ 1991, \aj, 102, 1091 

\bibitem[Patterson et al.(2001)]{patterson} Patterson, R. J., et al. 2001, 
in IAU Colloq. 183, Small Telescope Astronomy on Global Scales, eds. W. P. 
Chen, C. Lemme, \& B. Paczyski (ASP Conf. Ser. 246; San Francisco: ASP), 246

\bibitem[Tody(1986)]{tody} Tody, D.\ 1986, \procspie, 627, 733 

\bibitem[U.~S.~Naval Observatory \& Royal Greenwich 
Observatory(2001)]{usno} U.~S.~Naval Observatory \& Royal 
Greenwich Observatory 2001, The Astronomical Almanac for the year 2003, 
Washington: U.S.~Government Printing Office (USGPO) and London: The 
Stationery Office, 2001.

\bibitem[e.g., Yanny et al.(2003)]{yanny} Yanny, B.~et al. 2003, \apj, 
588, 824 

\end{thebibliography}
\end{document}